\documentclass[conference,9pt]{IEEEtran}
\usepackage[T1]{fontenc}
\usepackage[utf8]{inputenc}
\usepackage{newtxtext,newtxmath}
\usepackage{booktabs,array,graphicx,tikz,listings,cite}
\usepackage{balance}
\usetikzlibrary{arrows.meta,positioning}
\usepackage[hidelinks]{hyperref}
\hypersetup{pdftitle={Verified Netlist-to-RTL Recovery with LLM Agents},pdfauthor={Ahmet Efe, Recep Bugra Uludag, Ulya R. Karpuzcu}}
\newcommand{\sasc}{\texttt{sasc}}
\definecolor{wingreen}{rgb}{0,0.45,0}
\newcommand{\win}[1]{\textbf{\textcolor{wingreen}{#1}}}
\usepackage{multirow}
\usepackage{pifont}
\usetikzlibrary{positioning}
\begin{document}
\title{Verified Netlist-to-RTL Recovery with LLM Agents}
\author{\IEEEauthorblockN{Ahmet Efe, Recep Bugra Uludag, and Ulya R. Karpuzcu}
\IEEEauthorblockA{University of Minnesota, Twin Cities, Minneapolis, MN, USA\\
\{efe00002, uluda002, ukarpuzc\}@umn.edu}}
\maketitle

\begin{abstract}
Gate-level netlists circulate among IP integrators and foundries and can be extracted from chips. Reverse engineering them into readable register-transfer-level (RTL) code would help an adversary understand and modify a design, but remains difficult because word-level operations are scattered across many gates. Large language models (LLMs) can write and optimize RTL, but often fail to recover it from a whole netlist. We propose LOCA (LOgic Cone Attack), an LLM-based attack that divides an anonymized netlist into logic cones, has parallel agents rewrite them, and assembles compact RTL. Each cone contains the complete next-state logic of a group of related registers, giving an agent a small, meaningful task. Of five partitioning methods tested, only grouped cones lead to recovery. LOCA keeps only RTL proven equivalent to the netlist and measures its quality by compression. Function-preserving obfuscation does not stop LOCA on the tested designs, whereas unknown keys or hidden cell functions prevent it from proving recovery of the original design. Across 21 model--design pairs, LOCA recovers 17, versus 6 with whole-netlist prompting. Per-run recovery across models rises from 5\% to 74\%, at a median cost below one dollar with a low-cost model.

\end{abstract}

\section{Introduction}
Hardware reverse engineering aims to recover a circuit description that an engineer can understand and modify~\cite{RESEC}. An adversary may hold a gate-level netlist as an IP integrator or a foundry, or extract one from a chip~\cite{SoK-TCHES26,Chip-RE-Survey}. Such a netlist already allows cloning, but readable register-transfer-level (RTL) code exposes the counters, state machines, and datapaths that an adversary can change or reuse~\cite{Chip-RE-Survey}. Recovering RTL remains difficult because word-level operations are scattered across many unnamed gates.

Large language models (LLMs) are promising candidates for this task because they already write and optimize RTL~\cite{VerilogEval,RTLLM,RTLRewriter}. Prior work prompts an LLM with the whole netlist~\cite{Gate-Breaker}. In our experiments, however, whole-netlist prompting does not reliably recover readable RTL: its success depends on the model and on the size and structure of the design, and even larger models fail on some small controllers (Section~\ref{sec:enables}). We therefore ask how an LLM can recover designs that it cannot recover from the whole netlist.

A natural approach is to split the netlist into smaller pieces and recover each separately. However, how the netlist is split matters. In our experiments, conventional graph partitions and register cuts did not lead to recovery, even though they reduced the size of each task. Only logic cones covering the complete next-state logic of groups of related registers succeeded (Section~\ref{sec:enables}). These cones give each LLM a complete next-state function to rewrite.

\begin{figure}[t]
\centering
\includegraphics[width=\columnwidth]{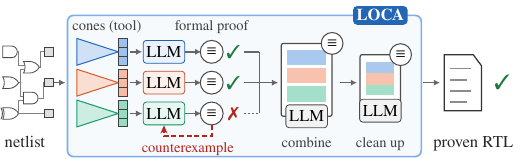}
\caption{LOCA cuts the netlist into cones; parallel LLM agents rewrite them, and a formal tool proves each rewrite ($\equiv$) or returns a counterexample. Agents combine and clean up the proven cones, with equivalence checked at each step.}
\label{fig:flow}
\end{figure}

LOCA (LOgic Cone Attack) assigns these cones to parallel LLM agents, each rewriting one cone using a small, low-cost model. A formal tool proves each rewrite or returns a counterexample, then verifies the assembled RTL against the netlist (Fig.~\ref{fig:flow}). The netlist serves as the specification, so no designer RTL is needed. We evaluate recovery using random simulation and formal equivalence. Yet equivalence alone is insufficient: a gate-level copy also passes the proof. We therefore introduce a compression measure, $R$, based on Boolean-operator reduction (Section~\ref{sec:measure}-A). We call a result a \emph{lift} when it is proven equivalent and meets the compression threshold.

We evaluate LOCA with Claude Haiku 5.5, Claude Sonnet 5.5, and GPT-6.1-sol on flattened, anonymized netlists in isolated sandboxes with opaque paths and no external file or web access. LOCA lifts 17/21 model--design pairs versus 6/21 for whole-netlist prompting, and succeeds in 88/119 runs versus 7/152 whole-netlist replies. The advantage persists against a whole-netlist Haiku agent given a comparable budget and the same checker (Section~\ref{sec:vsprompt}). On five fresh designs, LOCA achieves higher overall recovery rates with all three models. Follow-up experiments suggest recovered RTL is easier for an LLM to understand and modify, which an attacker needs to reuse or change a design (Section~\ref{sec:comprehension}).

Our contributions are:
{\setlength{\leftmargini}{1.2em}\begin{itemize}
\setlength{\itemsep}{0pt}\setlength{\parskip}{0pt}
\item We introduce LOCA, an LLM-based attack methodology that recovers compact, formally equivalent RTL from anonymized gate-level netlists without designer RTL.
\item We show that register-group logic cones outperform whole-netlist prompting, graph partitioning and clustering, and register cuts in our experiments.
\item We propose a reference-free recovery criterion combining formal equivalence with compression measure $R$, distinguishing recovered RTL from gate-level copies.
\item We evaluate three classes of netlist defenses: in our tests, function-preserving obfuscation does not prevent recovery, XOR/XNOR keys and hidden cell functions block recovery of the original design, and an LLM correctly guesses unlock sequences.
\end{itemize}}


\section{Threat Model and Related Work}
\label{sec:threat}

\subsection{Attacker and goal}
The attacker is an IP integrator, a foundry, or anyone who extracts a netlist from a chip. It holds a flattened gate-level netlist but no RTL, specification, or testbench, and cannot query a working chip (the oracle-less setting). Using commercial LLMs and open-source synthesis and formal tools, it aims to recover RTL that it can understand, reuse, or modify, for example to insert a hardware Trojan~\cite{HAL}. A foundry that can modify a layout before fabrication is one such attacker~\cite{A2}. For this goal, recovered RTL must be equivalent to the netlist and substantially more compact (Section~\ref{sec:measure}).

\subsection{Requirements and defenses}
LOCA can prove recovered RTL against a netlist with known gate functions. Function-preserving transformations, such as resynthesis and retiming, change the netlist's structure but not its behavior (Section~\mbox{\ref{sec:defense}-A}). XOR/XNOR locking adds key inputs: LOCA can recover proven RTL of the locked circuit, but recovering the original design requires the correct key. Hidden cell functions, a netlist-level model of camouflaging~\cite{Camo-CCS13,Decamouflage}, obscure what gates compute, for example by labeling an AND gate as OR. Existing locking attacks query a working chip~\cite{SAT-Attack,LLM-Unlock} or infer keys from circuit structure~\cite{FALL,SAIL-AsianHOST18,GNNUnlock,SCOPE,OMLA}; LOCA's lifting procedure does neither. We separately test whether LLMs can infer keys from locked netlists or lifted RTL (Section~\ref{sec:defense}). Without a working chip or the original design, the attacker cannot confirm a guessed XOR/XNOR key.

\subsection{Related work}
Table~\ref{tab:checks} compares prior methods. Classical netlist analysis recovers registers, words, operators, functional blocks, and state machines~\cite{DANA,WordRev,Subramanyan-TETC14,RELIC-FUN,REFSM}, and learned methods classify sub-circuits, state registers, arithmetic blocks, or word-level groupings~\cite{GNN-RE,ReIGNN,Gamora,ReBERT}; these label parts of the netlist but do not produce complete RTL. HAL supports analyst-driven reverse engineering~\cite{HAL}; LOCA runs without an analyst. RERTL recovers state-machine-centered RTL using a file that identifies register words~\cite{RERTL}; LOCA needs no such file and also recovers datapath operations such as arithmetic and shifts. LLMs generate RTL from specifications~\cite{VerilogEval,RTLLM}, fix hardware security bugs~\cite{LLM-SecFix}, and rewrite RTL with formal feedback~\cite{RTLRewriter}, but start from a specification or RTL, not a netlist.

\begin{table}[t]
\caption{Netlist-recovery methods: validation and evaluation settings.}
\label{tab:checks}
\centering\scriptsize
\setlength{\tabcolsep}{3pt}
\renewcommand{\arraystretch}{0.85}
\newcommand{\tyes}{{\color{green!50!black}\ding{51}}}
\newcommand{\tno}{{\color{red!65!black}\ding{55}}}
\begin{tabular}{llccccc}
\toprule
 & Recovers
 & \shortstack{Simu-\\lation}
 & \shortstack{Formal\\equiv.}
 & \shortstack{Ref.-free\\measure}
 & \shortstack{Names\\hidden}
 & \shortstack{Synth.\\variants}\\
\midrule
Classical~\cite{DANA,WordRev,Subramanyan-TETC14,RELIC-FUN,REFSM}
 & Words, ops., FSMs & -- & -- & -- & -- & --\\
RERTL~\cite{RERTL}
 & FSM-based RTL & -- & -- & -- & -- & --\\
Gate-Breaker~\cite{Gate-Breaker}
 & Full RTL & \tyes & \tno & \tno & \tno$^{a}$ & \tno\\
\textbf{LOCA (ours)}
 & \textbf{Full RTL} & \tyes & \tyes & \tyes & \tyes$^{b}$ & \tyes\\
\bottomrule
\end{tabular}
\par\smallskip
{\scriptsize --: not compared. $^{a}$Wire names hidden, port names kept. $^{b}$All names hidden, incl.\ ports.}
\end{table}

The closest work, Gate-Breaker, gives an LLM the whole netlist at once, accepts answers that pass simulation against a reference testbench, and scores them by abstract-syntax-tree and control-flow-graph similarity to the original RTL~\cite{Gate-Breaker}; our attacker has neither a testbench nor the designer's RTL. LOCA instead investigates how LLMs can reconstruct higher-level operations and next-state behavior directly from anonymized gates, using register-group cones without predefined operator-recovery templates. Section~\ref{sec:vsprompt} compares LOCA with reconstructed Gate-Breaker prompts under the same checks.


\section{LOCA Methodology}
\label{sec:attack}
Whole-netlist rewriting is limited by design size and structure, and the limit depends on the model (Section~\ref{sec:enables}). LOCA instead extracts logic cones: the fan-in logic of register inputs or primary outputs, traced back to register outputs and primary inputs. Logic cones are standard in equivalence checking; our contribution is to use them for LLM-based RTL reverse engineering. These cones give LLM agents well-defined rewriting tasks rather than arbitrary graph fragments. Agents work in parallel, while formal tools verify each rewrite and every assembly step.

LOCA takes a flattened netlist with known gate functions and produces equivalent RTL in three stages (Fig.~\ref{fig:flow}): (A)~deterministic tools extract register-group cones; (B)~parallel LLM agents rewrite the cones, with a checker proving each accepted rewrite; and (C)~more agents assemble and simplify the proven results, while a script selects the proven version with the fewest operators. Fig.~\ref{fig:example} shows excerpts of one recovery. Cone checks compare rewrites against the original cones, assembly checks compare successive versions, and the final proof compares the result against the input netlist. No designer RTL or testbench is needed. LLMs reconstruct higher-level logic without predefined operator templates; tools built on Yosys and ABC handle extraction, grouping, simulation, verification, and final selection~\cite{Yosys,ABC}. Failed rewrites are retried or discarded, leaving the original cone logic in place, so a wrong answer never enters the result.

\begin{figure}[t]
\centering
\begin{minipage}[t]{0.49\columnwidth}
\textbf{Input netlist (excerpt)}
\begin{lstlisting}[basicstyle=\ttfamily\fontsize{6.5}{7.6}\selectfont,frame=single,framesep=2pt,rulecolor=\color{black!45},xleftmargin=2.4pt,xrightmargin=2.4pt]
assign wfea = w14f & ~(we11);
assign w664 = w0ec & we11;
assign wf0d = wfea | w664;
assign wbca = wc7f & ~(we11);
assign w4e1 = wcaf & we11;
assign wb44 = wbca | w4e1;
assign w136 = wc86 & ~(we11);
assign w8c8 = w758 & we11;
assign w408 = w136 | w8c8;
\end{lstlisting}
\end{minipage}\hfill
\begin{minipage}[t]{0.49\columnwidth}
\textbf{Recovered RTL (excerpt)}
\begin{lstlisting}[basicstyle=\ttfamily\fontsize{6.5}{7.6}\selectfont,frame=single,framesep=2pt,rulecolor=\color{black!45},xleftmargin=2.4pt,xrightmargin=2.4pt]
if (x43b) begin
  lanes[8*phase +: 8] <=
    {xbe5, xf02, x7e3, xc2f,
     x343, x7fa, x06b, x6ef};
  phase <= phase + 2'd1;
end
if (shift_en)
  chain <= {chain[6:0],
            xin[0]};
\end{lstlisting}
\end{minipage}

\caption{Separate excerpts from an anonymized netlist and its proven RTL recovery, not a line-by-line translation. Readable names were chosen by the model; names such as \texttt{w4e1} and \texttt{x43b} are anonymized.}
\vspace{-0.1cm}

\label{fig:example}
\end{figure}

\subsection{Cone extraction and register grouping}
LOCA first extracts a cone for each flip-flop input and primary output, tracing its logic back to register outputs and primary inputs. Cones may overlap: a gate feeding several registers can appear in multiple cones, and each agent receives its own copy. We keep these copies because merging overlapping cones can quickly exceed the gate count an LLM can handle, preventing recovery. Related flip-flops are then grouped into cones containing their complete next-state logic, as described below. After rewriting, hundreds of gates may become a few lines of RTL, making duplicated logic easier to remove during assembly (Section~\mbox{\ref{sec:attack}-C}). LOCA targets single-clock synchronous netlists without assuming a particular architecture, such as a state machine or datapath.

LOCA uses standard structural analysis to group related flip-flops from the anonymized netlist~\cite{DANA,WordRev}, such as bits of a multi-bit register or a one-hot state encoding. It repeatedly separates flip-flops with different enable signals, input/output connections, or dependencies on other register groups until the grouping stabilizes. Each group then forms a cone containing its complete next-state logic. Cones exceeding 600 gates are split into smaller tasks while keeping structurally related bit slices, such as adder carry logic, together.

\subsection{Per-cone rewriting and equivalence checking}
Each cone, or part of a split cone, becomes a small gate-level Verilog module for its own LLM agent, with the registers it reads as inputs. No agent sees the whole netlist. The prompt asks for compact RTL rather than a gate-by-gate copy. A formal checker, not the agent, proves each rewrite equivalent to its original cone. Normally this is a combinational check: the registers a cone reads are treated as free inputs on both sides, keeping the proof local to the cone. If a rewrite is correct only for reachable register states, such as a one-hot encoding, LOCA instead uses sequential equivalence.

Assembly, clean-up, and final checks are also sequential because they may change the registers. Sequential proofs limit design size: on our designs, the two final proofs of a run take 2\,s together at the median and at most 251\,s, each within a 600\,s limit. Unlike random simulation, formal equivalence covers rarely exercised behavior. If simulation or proof fails, the agent receives a counterexample and can retry up to twice. Failed rewrites leave gates intact, which reduces compression but never affects correctness.

\subsection{Assembly and whole-design proof}
Assembly starts from the proven cones, which may contain duplicated logic. Agents using the same LLM group the cones into blocks and rewrite each block as one module, seeing only the proven RTL rather than the original gates. Next, for at most eight rounds, an agent examines the modules and their connections and either merges modules or stops. Finally, clean-up agents attempt to remove repeated logic across the design. Each step must pass sequential equivalence against the preceding version; otherwise, the change is rejected.

Every proven version is a candidate. A script selects the candidate with the fewest operators, including arithmetic and comparisons, breaking ties by line count. The selected RTL must pass whole-design sequential equivalence against both the converted netlist used by LOCA and the original input netlist. A final co-simulation tests practical portability by checking that the recovered Verilog compiles, runs, and matches the netlist in a standard simulator, separately from the earlier random simulation of Fig.~\ref{fig:checks}. Both proofs and the co-simulation must pass.


\section{Evaluation Method}
\label{sec:measure}
\subsection{Evaluation criteria}

\begin{figure}[t]
\centering
\begin{tikzpicture}[font=\footnotesize, >=stealth, node distance=3.2mm,
  box/.style={draw, rounded corners=2pt, align=center, minimum height=9mm, text width=17mm, inner sep=1.5pt},
  ok/.style={font=\scriptsize, align=left, anchor=north west}]
\node[box] (s) {\textbf{Functional}\\\textbf{Simulation}};
\node[align=center, font=\footnotesize\bfseries, left=of s, yshift=-1.6mm] (a) {LLM\\Answer\\{\color{orange!85!black}?}};
\node[box, right=of s] (p) {\textbf{Formal}\\\textbf{Verification}};
\node[box, right=of p] (c) {\textbf{Compression}\\$(R)$};
\node[right=3.2mm of c, align=center, font=\footnotesize\bfseries, yshift=-1.6mm] (l) {Lifted\\{\color{green!50!black}\ding{51}}};
\draw[->] (a.east |- s.west) -- (s.west);
\draw[->] (s) -- (p);
\draw[->] (p) -- (c);
\draw[->] (c.east) -- (l.west |- c.east);
\newcommand{\yes}{{\color{green!50!black}\ding{51}}}
\newcommand{\no}{{\color{red!65!black}\ding{55}}}
\node[ok] at ([yshift=-1mm]s.south west) {\yes\ finds bugs fast\\\no\ no proof};
\node[ok] at ([yshift=-1mm]p.south west) {\yes\ exact proof\\\no\ no quality check};
\node[ok] at ([yshift=-1mm]c.south west) {\yes\ checks RTL quality\\\yes\ no designer RTL};
\end{tikzpicture}
\caption{Reference-free lifting checks: simulation detects errors, formal verification proves equivalence, and compression measures abstraction without designer RTL. Stages show logical order; simulation and proof run in parallel.}
\label{fig:checks}
\end{figure}
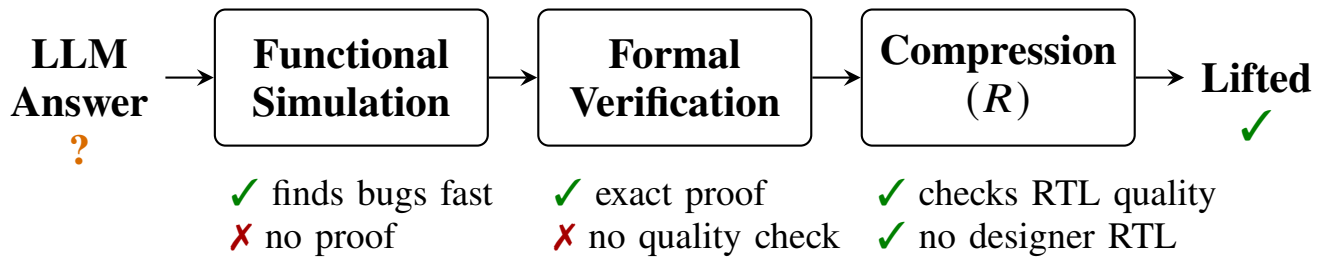

We evaluate recovered RTL along two dimensions: whether it preserves the netlist's behavior and whether it replaces gate-level logic with higher-level operations. Our checks require neither the designer's RTL nor a reference testbench (Fig.~\ref{fig:checks}).

\emph{(1) Does it have obvious mistakes?} Random simulation quickly detects mismatches between the recovered RTL and the netlist, rejecting incorrect answers. Passing simulation, however, does not prove equivalence, because rare input sequences may never be exercised.

\emph{(2) Is it formally equivalent to the netlist?} After one reset cycle, the recovered RTL and netlist must produce identical outputs in every subsequent cycle, under any input sequence and later resets~\cite{SEC-ICCD06,SigCorr-ICCAD08}. Equivalence alone does not measure abstraction: a gate-level copy also passes the proof.

\emph{(3) Is it more abstract than the netlist?} Beyond functional correctness, recovered RTL should express higher-level operations rather than reproduce individual gates. We need to measure this abstraction without relying on the designer's RTL, which the attacker does not have, or on a reference testbench.

For example, a counter written as \texttt{counter + 1} replaces a chain of XOR and AND gates. Line count is unreliable because long expressions can fit on one line, while RTL similarity~\cite{Gate-Breaker} requires a reference. We therefore define a compression measure, $R$, based on the reduction in bitwise Boolean operators. Our netlists contain only AND, OR, XOR, their inverted forms, and NOT, so we count occurrences of \texttt{\&}, \texttt{|}, \texttt{\^{}}, and \texttt{\~{}} in both the answer and netlist:
\begin{equation}
R = 1 -
\frac{N_{\mathrm{bool}}(\mathrm{answer})}
     {N_{\mathrm{bool}}(\mathrm{netlist})}.
\label{eq:compression}
\end{equation}
A gate-level copy has $R=0$, while RTL without these operators has $R=100\%$. Case tables and arithmetic can inflate $R$, so Section~\ref{sec:robust} also counts all operators and bounds table entries. The models never see $R$ or the threshold, so they cannot deliberately exploit the metric. Their prompts ask for readable, compact RTL.

We call a run \emph{lifted} when its RTL is formally equivalent to the netlist and removes more than 60\% of its Boolean operators ($R>60\%$). We fixed this threshold before the main runs, so that a lift removes well over half of the operators. The exact threshold is not critical: LOCA remains ahead of whole-netlist prompting from 30\% to 80\% (Section~\ref{sec:robust}). We also examine how $R$ relates to LLM comprehension (Section~\ref{sec:comprehension}).

\subsection{Experimental setup and validation}
\textbf{Designs.} We use ITC'99 controllers~\cite{ITC99}, ranging from small state machines (b01 compares two serial streams) to a processor subset (b15); IWLS IP cores~\cite{IWLS05}; generated sequential circuits, including a 12-bit up/down counter (\texttt{updown}), a 16-entry FIFO (\texttt{fifo16x8}), a UART receiver (\texttt{uartrx16}), and a pipelined multiply-accumulate unit (\texttt{macpipe8}); and three \emph{functional mutants}: b05m (new data table, state encoding, and display font), a \sasc{} mutant, and an \texttt{ss\_pcm} mutant. These mutants modify public designs so they are no longer equivalent to their originals (Section~\ref{sec:robust}).

Yosys and ABC~\cite{Yosys,ABC} synthesize, optimize, flatten, anonymize (including ports), and bit-blast each design into two-input gates. Multiplexers, adders, and other higher-level cells disappear, leaving the model to recover words and operators from basic gates alone. Each netlist is formally verified against its source after one reset cycle. All designs use one clock and known asynchronous reset values.

We developed LOCA's prompts using ITC'99, b05m, \texttt{i2c}, \texttt{serv} (a bit-serial RISC-V core), and our generated circuits. The IWLS cores \sasc{} (a UART-like serial controller), \texttt{simple\_spi} and \texttt{spi} (SPI masters), and \texttt{systemcdes} (a DES cipher) were not used for prompt development. Before the final-version runs, we declared five \emph{fresh designs}: three unused public cores, \texttt{ss\_pcm} (a PCM audio interface, 520 gates), \texttt{usb\_phy} (a USB~1.1 PHY, 656), and \texttt{sockit\_owm} (a 1-Wire bus master, 629), plus two unpublished designs, a byte-command decoder (610) and a request policer (634). On acceptance, we will release the generated, mutated, and unpublished designs for reproduction.

We evaluate Claude Haiku 5.5 (Haiku), Claude Sonnet 5.5 (Sonnet), and GPT-6.1-sol (GPT), with reasoning effort set to ``high''. The counted LOCA runs used Claude Code 2.1.311--2.1.317 and Codex CLI 0.160.1. We also report three exploratory Gemini 3.1 Pro (Gemini) runs through Cursor.

\textbf{Isolation.} Frontier LLMs may recognize designs from module names or file paths, search for known RTL online, or use auxiliary tools to bypass the intended recovery task. To prevent this, each agent runs in an isolated sandbox with opaque paths, no web access, and only the prescribed LOCA tools permitted. A guard rejects prompts or paths naming benchmarks and stops runs when generated block names match guarded patterns (a \emph{guard stop}). Every transcript is audited. Of over 23,000 screened calls, 29 were flagged; an extra check of these found no access to files outside their own task folders.

\textbf{Counting.} Headline results include only the 21 primary model--design pairs (unmarked rows of Table~\ref{tab:pairs}). For each method, a pair succeeds if at least one attempt meets the lifting criterion. A LOCA \emph{run} is one execution of the pipeline, a whole-netlist \emph{reply} is one answer to a prompt, and a \emph{session} is one conversation with the whole-netlist agent of Table~\ref{tab:gb}. Non-answers and guard stops count as failures and stay in every denominator.

\textbf{Cross-check of the proofs.} Every counted lift passes the main equivalence checker. We independently rechecked a set of 152 LOCA lifts and 35 whole-netlist lifts using k-induction in SymbiYosys~\cite{SymbiYosys} and IC3~\cite{IC3}, with invariants checked by Z3~\cite{Z3}. These checks are separate from the main checker, which uses ABC's property-directed reachability~\cite{PDR} among its engines. They confirm 147 of the 152 LOCA lifts and all 35 whole-netlist lifts, with no errors found; the remaining five exceed the time limit. The 27 Sonnet and GPT lifts on fresh designs rely on the main proof, as do lifts from later experiments, such as the defense probes and the mutant controls.


\section{Experimental Evaluation}
\label{sec:results}

\subsection{LOCA versus whole-netlist prompting}
\label{sec:vsprompt}

\textbf{Main result.} Across 21 primary model--design pairs (Table~\ref{tab:pairs}), LOCA lifts 17/21, compared with 6/21 for whole-netlist prompting, including Gate-Breaker prompts. The standard prompt lifts five pairs; Gate-Breaker adds Haiku on \texttt{updown}. Across repeated attempts, LOCA succeeds in 88/119 runs versus 7/152 standard whole-netlist replies, a roughly 16$\times$ higher success rate; per model, it is about 80, 7.5, and 4 times higher for Haiku, GPT, and Sonnet. On b10, no whole-netlist reply succeeds (0/22, including Gemini), while GPT with LOCA lifts 3/3 runs. On b09, Haiku with LOCA lifts 8/10 runs, while all 19 whole-netlist replies from Haiku, Sonnet, and GPT fail.

\begin{table}[t]
\caption{Whole-netlist prompting versus LOCA: lifts/attempts, best proven $R$ (\%), and median LOCA model cost. Prompting counts standard-prompt replies; LOCA counts runs. $^{\star}$: earlier LOCA version, excluded from primary counts. $^{\dagger}$: declarations reordered; proofs and co-simulation repeated. ---: no proven answer. Highlighted entries indicate higher lift frequency.}
\label{tab:pairs}
\centering\scriptsize
\setlength{\tabcolsep}{2.5pt}
\renewcommand{\arraystretch}{0.85}
\begin{tabular}{llrrrrrr}
\toprule
Model & Design & Gates & Prompt. & \shortstack[r]{\textbf{LOCA}\\\textbf{(ours)}} & \shortstack[r]{Prompt.\\$R$ (\%)} & \shortstack[r]{\textbf{LOCA}\\$R$ (\%)} & \shortstack[r]{LOCA\\cost (\$)}\\
\midrule
Haiku 5.5 & b01 & 43 & 1/13 & \win{9/10} & 72 & \win{95} & 0.09\\
 & b09 & 105 & 0/13 & \win{8/10} & 57 & \win{92} & 0.41\\
 & updown & 110 & 0/13 & \win{10/10} & 44 & \win{100} & 0.04\\
 & b10 & 111 & 0/13 & 0/10 & 14 & 27 & 0.28\\
 & b07 & 219 & 0/13 & \win{6/10} & 6 & \win{100} & 0.50\\
 & b04 & 312 & 0/13 & \win{10/10} & 8 & \win{98} & 0.45\\
 & fifo16x8 & 486 & 0/13 & \win{10/10} & 1 & \win{99} & 0.28\\
 & sasc & 882 & 0/13 & \win{9/10} & 38 & \win{96} & 0.71\\
\midrule
Sonnet 5.5 & b01$^{\star}$ & 43 & 0/3 & \win{2/2} & 29 & \win{91} & 0.26\\
 & b09 & 105 & 0/3 & \win{1/3} & 48 & \win{71} & 1.49\\
 & b10 & 111 & 0/3 & 0/3 & 4 & 26 & 1.42\\
 & b04$^{\star}$ & 312 & 0/4 & \win{4/4} & 0 & \win{97} & 1.20\\
 & b05m & 850 & 0/5 & \win{3/3} & 0 & \win{82} & 6.90\\
 & sasc & 882 & 2/3 & \win{3/3} & 93 & \win{94} & 3.01\\
 & b05 & 890 & 0/7 & \win{1/3} & 0 & \win{66} & 6.04\\
 & simple\_spi & 1151 & 2/4 & \win{3/3} & 89 & \win{92} & 5.80\\
 & b11 & 1235 & 0/5 & 0/3 & 0 & 45 & 8.34\\
 & systemcdes & 3375 & 0/3 & 0/3 & 0 & 39 & 5.15\\
\midrule
GPT-6.1-sol & b09 & 105 & 0/3 & \win{3/3} & 56 & \win{90} & 0.64\\
 & b10 & 111 & 0/3 & \win{3/3} & 36 & \win{96} & 1.03\\
 & b13 & 269 & 1/3 & \win{3/3} & 68 & \win{80} & 1.09\\
 & b12 & 1093 & 1/3 & \win{3/3} & 75 & \win{90} & 2.36\\
 & b11$^{\star}$ & 1235 & 1/5 & \win{1/1} & 96 & \win{96} & 1.71\\
 & serv & 1648 & 0/3 & \win{3/3} & --- & \win{74} & 5.89\\
 & b15$^{\star\dagger}$ & 20794 & 0/3 & \win{1/2} & --- & \win{73} & 25.22\\
\midrule
Gemini 3.1 Pro & b10$^{\star}$ & 111 & 0/3 & \win{3/3} & 0 & \win{96} & n/a\\
\bottomrule
\end{tabular}
\end{table}

The final LOCA version lifts \texttt{serv} (1,648 gates) in 3 of 3 GPT runs and \texttt{spi} (4,909 gates) in 1 of 4 Sonnet runs, its largest lift (Table~\ref{tab:reach}). An earlier version lifted b15$^{\star}$, a 20,794-gate Intel 80386 subset ($R=73\%$, 1 of 2 runs), after a scripted reordering of its declarations; all 3 whole-netlist replies on b15 returned no RTL.

\begin{table}[t]

\caption{Additional LOCA runs: lifts/runs and best lifted $R$ (\%). $^{\star}$: earlier LOCA version, excluded from primary counts. $^{\dagger}$: proven RTL co-simulated after automatic syntax correction. ---: no lift.}

\label{tab:reach}
\centering\scriptsize\setlength{\tabcolsep}{1.6pt}\renewcommand{\arraystretch}{0.87}
\begin{tabular}{llrrr|llrrr}
\toprule
Model & Design & Gates & Lifted & $R$ & Model & Design & Gates & Lifted & $R$\\
\midrule
Sonnet & b02$^{\star}$ & 23 & 2/2 & 100 & Sonnet & fifo16x8 & 486 & 3/3 & 99\\
Sonnet & b06$^{\star}$ & 42 & 2/2 & 92 & Sonnet & macpipe8$^{\star}$ & 570 & 1/1 & 100\\
Sonnet & spimaster$^{\star}$ & 80 & 1/1 & 84 & GPT & sasc & 882 & 3/3 & 96\\
Sonnet & b03$^{\star}$ & 90 & 3/3 & 82 & Haiku & sasc mutant & 882 & 3/3 & 92\\
Sonnet & b08$^{\star}$ & 138 & 2/2 & 96 & Haiku & b12 & 1093 & 0/1 & ---\\
Sonnet & b07 & 219 & 1/3 & 85 & Sonnet & b12$^{\star}$ & 1093 & 3/3 & 78\\
Sonnet & uartrx16$^{\star}$ & 258 & 1/1 & 91 & Haiku & b11 & 1235 & 0/1 & ---\\
Haiku & b13 & 269 & 1/1 & 72 & Sonnet & b11$^{\star}$ & 1235 & 4/4 & 96\\
Sonnet & b13$^{\star}$ & 269 & 1/2 & 68 & Sonnet & spi$^{\dagger}$ & 4909 & 1/4 & 87\\
\bottomrule
\end{tabular}
\end{table}

\textbf{Fresh designs.} On five designs held out from development, Haiku with LOCA lifts all five and succeeds in 20/25 runs, versus 20/50 whole-netlist replies and 1/50 Gate-Breaker replies. Of the 20 successful whole-netlist replies, 17 come from \texttt{ss\_pcm} and \texttt{sockit\_owm}; only LOCA lifts \texttt{usb\_phy} and the unpublished request policer with Haiku. The advantage is smaller with stronger models but remains: Sonnet lifts 13/15 LOCA runs versus 36/50 whole-netlist replies, and GPT lifts 14/15 versus 34/50. LOCA has a higher lift rate on three of the five designs with each model.

\textbf{Baselines.} On the seven designs in Table~\ref{tab:gb}, LOCA lifts 62/70 Haiku runs, compared with 1/91 replies using our standard prompt and 1/75 using the Gate-Breaker prompt. Most proven Gate-Breaker replies remain close to gate level: 43 of 46 remove less than half of the netlist's Boolean operators. To account for LOCA's multiple agent calls, we give a whole-netlist Haiku agent the same checker and a LOCA run's median cost as its budget. It lifts only 3/21 sessions, stopping well below that budget on three designs.

\begin{table}[t]
\caption{Haiku baselines on seven designs lifted by at least one method. Lifted: successes/attempts; mean $R$ counts failed checks as zero. Gate-Breaker uses reconstructed prompts. Agent: a whole-netlist agent with the same checker and a budget equal to the median LOCA run cost (3 sessions per design). Bold: highest lift rate or mean $R$.}
\label{tab:gb}
\centering\scriptsize
\setlength{\tabcolsep}{2pt}
\renewcommand{\arraystretch}{0.87}
\begin{tabular}{llrrrrrrr}
\toprule
 & & b01 & b09 & updown & b07 & b04 & fifo16x8 & sasc\\
 & Gates & 43 & 105 & 110 & 219 & 312 & 486 & 882\\
\midrule
\multirow{4}{*}{Lifted}
 & Prompt. & 1/13 & 0/13 & 0/13 & 0/13 & 0/13 & 0/13 & 0/13\\
 & Gate-Breaker~\cite{Gate-Breaker} & 0/12 & 0/11 & 1/11 & 0/10 & 0/10 & 0/10 & 0/11\\
 & Agent & 0/3 & 0/3 & 1/3 & 0/3 & 0/3 & 1/3 & 1/3\\
 & \win{LOCA (ours)} & \win{9/10} & \win{8/10} & \win{10/10} & \win{6/10} & \win{10/10} & \win{10/10} & \win{9/10}\\
\midrule
\multirow{4}{*}{\shortstack[l]{Mean\\$R$ (\%)}}
 & Prompt. & 33 & 37 & 12 & 1 & 1 & 0 & 3\\
 & Gate-Breaker~\cite{Gate-Breaker} & 24 & 0 & 34 & 0 & 0 & 0 & 0\\
 & Agent & 26 & 48 & 60 & 8 & 4 & 33 & 52\\
 & \win{LOCA (ours)} & \win{79} & \win{71} & \win{100} & \win{59} & \win{88} & \win{98} & \win{86}\\
\bottomrule
\end{tabular}
\end{table}

\subsection{Ablation: decomposition and feedback}
\label{sec:enables}
\textbf{Size and structure limit.} Whole-netlist prompting depends on both the model and the circuit's size and structure. On a ladder of increasingly large generated and ITC'99 designs, Haiku failed to lift any design with 24 or more flip-flops. The ladder excludes the fresh designs, some of which Haiku lifts from the whole netlist (Section~\ref{sec:vsprompt}). Sonnet and GPT recovered some designs with 900--1,200 gates but failed on much smaller ITC'99 controllers (Table~\ref{tab:pairs}). To test the effect of structure, we changed the register encoding of \texttt{ss\_pcm} without changing its behavior. Haiku's whole-netlist success dropped from 10/10 to 1/10 replies, while LOCA still lifted 5/5 runs.

\begin{figure}[t]
\centering
\includegraphics[width=0.9\columnwidth]{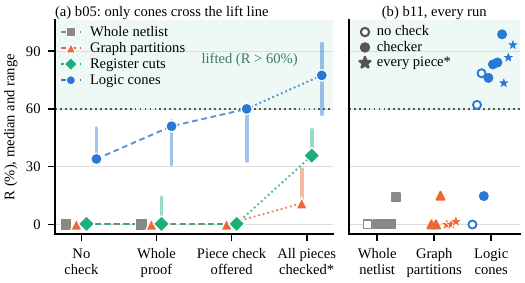}
\caption{Sonnet ablation of decomposition and per-piece verification. (a) b05: median and range of proven $R$. (b) b11: each marker represents one proven answer. * denotes mandatory per-piece proofs; failed proofs are omitted.}
\label{fig:ladder}
\end{figure}

\textbf{Decomposition.} We test whether LOCA benefits simply from smaller tasks or from where the netlist is cut. Fig.~\ref{fig:ladder} compares whole-netlist prompting, balanced $k$-way hypergraph partitions (Mt-KaHyPar~\cite{MtKaHyPar}), register cuts, and grouped cones using Sonnet on b05 and b11 (register cuts on b05 only). The balanced partitions have sizes similar to the cones. We also test unbalanced min-cut partitions (Mt-KaHyPar, imbalance 0.30) and community detection (Leiden~\cite{Leiden}), reported in the text. Graph partitions can cut through functional boundaries, while register cuts preserve each flip-flop's next-state logic without grouping related bits.

None of the four alternative partitioning methods leads to recovery in our tests. Balanced partitions remain near gate-level copies even with feedback. With every piece checked, unbalanced and Leiden partitions lift 0/12 runs (three per method and design; all proven, $R$ at most 30\%). Register cuts on b05 reach only $R=33$--$49\%$, below the lifting threshold. Even providing register-dependency hints alongside the whole netlist reaches only $R=47\%$ on b05. We attribute the advantage of grouped cones to preserving complete next-state logic of related register bits rather than reducing task size.

\textbf{Effect of per-piece proofs.} We first tested whether giving an LLM logic cones without checking each rewrite was enough. In a single unchecked reply, Haiku lifted only 2 of 5 designs (b01, b09, \texttt{updown}, b10, and \sasc{}), compared with 4 using LOCA. Sonnet lifted only 1 of 9 small designs. When we required a formal proof for every cone, Sonnet lifted 5 of 6 runs on b05 and all 3 runs on b11. In contrast, none of the graph-partition runs succeeded, despite receiving a similar number of checks (Fig.~\ref{fig:ladder}). Unlike LOCA's parallel agents, this ablation processes cones sequentially, avoiding duplicated shared logic. In the Haiku and Sonnet runs on b10 and b11, duplicated logic survived clean-up and kept the result below the lifting threshold. An earlier LOCA version, with a single-call clean-up, lifted b11 with Sonnet in 4 of 4 runs (Table~\ref{tab:reach}); in the final LOCA version, the iterative clean-up stops early (0 of 3).

\subsection{Robustness and memorization}
\label{sec:robust}

\textbf{Threshold and operator count.} We test whether LOCA's advantage depends on the chosen lifting threshold or the operators counted. At every threshold from 30\% to 80\%, LOCA alone lifts 8--11 of the 21 primary model--design pairs. When counting all operators, including selections, comparisons, arithmetic, and flip-flop assignments, LOCA still lifts 17 pairs versus 8 for whole-netlist prompting. We check whether recovered RTL hides logic in constructs not counted by $R$. Across proven LOCA answers, \texttt{case} entries, \texttt{?:} selections, and table lookups account for at most 9\% of the netlist's gates.

\textbf{Synthesis tools and standard cells.} We test whether conventional synthesis tools can produce comparably compact RTL without LLMs. On \sasc{}, \texttt{simple\_spi}, b05, b05m, and \texttt{spi}, the smallest proven RTL produced by Yosys and ABC contains 2.3--9.9 times as many operators as the designer's RTL. Each measured LOCA lift has at most 39\% of the operators of the tool-only result. We also test \texttt{ss\_pcm} and b04 mapped to ASAP7 standard cells~\cite{ASAP7}. After reducing these cells to generic gates, Haiku with LOCA lifts all 6 runs, as it does on the generic-gate netlists.

\textbf{Memorization.} A model might recover a public design by recalling its RTL rather than interpreting the netlist. To test this possibility, we modified the \sasc{} serial protocol in four places, each sufficient to break equivalence with the public core. Sonnet lifted the original in 4 of 5 whole-netlist replies but returned gate-level copies of the modified design in all 5. Haiku with LOCA lifted the modified design in 3 of 3 runs. GPT lifted both versions through whole-netlist prompting but failed on \texttt{pci\_spoci}, a 908-gate public core not used in development, which LOCA lifted ($R=66\%$). Haiku also lifts the public core \texttt{ss\_pcm} from the whole netlist. A mutant with four behavior changes gives the same result (10 of 10 replies; LOCA 5 of 5), so Haiku does not recall the public RTL. Its whole-netlist success depends instead on repeated register structure: re-encoding the registers drops it to 1 of 10 replies (Section~\ref{sec:enables}), while LOCA lifts every variant in 5 of 5 runs.

\subsection{LLM comprehension and editing of lifted RTL}
\label{sec:comprehension}
\begin{figure}[t]
\centering
\includegraphics[width=0.9\columnwidth]{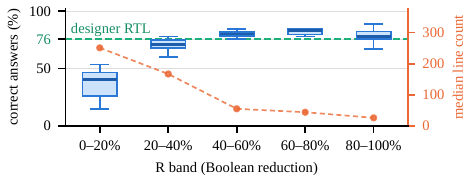}
\caption{Haiku's comprehension of b09 across 20-point $R$ bands. Boxes show mean accuracy; orange shows median line count (right axis). Dashed lines mark the designer's RTL score and the netlist's 279 lines. Results use 15 of 19 questions; four questions answered correctly in at least 90\% of sessions were excluded after initial testing.}
\label{fig:comprehension}
\end{figure}

Formal equivalence establishes correctness, and compression measures how much gate-level logic was replaced, but neither shows whether recovered RTL is easier to use. To test comprehension, we ask Haiku 19 predefined behavior questions about 42 proven, anonymized versions of one design, b09, ranging from the netlist to the designer's RTL. Answers are checked by simulation. In this single-design, single-model experiment, accuracy rises from 44\% below $R=20\%$ to 82--87\% from $R=40\%$ onward, compared with 79\% for the designer's RTL. Fig.~\ref{fig:comprehension} shows a 15-question subset selected after the initial sessions by removing four questions answered correctly in at least 90\% of sessions. On this subset, accuracy rises from 41\% to 78--83\%, versus 76\% for the designer's RTL.

We also test whether lifted RTL is easier to modify. For \texttt{ss\_pcm} and \texttt{sockit\_owm}, we ask Haiku to make the same small change starting from either lifted RTL or the gate-level netlist. An edit succeeds only if it is formally equivalent to the modified designer RTL. Edits to lifted RTL succeed in 10 of 10 attempts, compared with 7 of 10 on the netlist.

\section{Evaluation of Netlist Defenses}
\label{sec:defense}
We evaluate three forms of netlist protection against LOCA: function-preserving obfuscation, logic locking including sequential (state-machine) obfuscation, and hidden cell functions. They differ in whether the attacker has enough information to recover the original design's function (Section~\mbox{\ref{sec:threat}-B}). Table~\ref{tab:defense} summarizes the main probes on b01, b09, and \sasc{}, which Haiku lifts in most unprotected runs, providing a baseline for comparison.

\subsection{Function-preserving obfuscation}
We test five function-preserving transformations (Table~\ref{tab:defense}). Haiku with LOCA lifts 35 of 39 obfuscated runs. The four failures occur on b01 (3) and b09 (1), including three guard stops counted as failures (Section~\mbox{\ref{sec:measure}-B}). For comparison, Haiku fails 1 of 10 unprotected b01 runs and 2 of 10 unprotected b09 runs (Table~\ref{tab:pairs}). At this sample size, the results show no clear reduction in LOCA's recovery rate due to obfuscation, since every transform keeps the netlist's function.

\subsection{Logic locking: reconstruction and key inference}
We insert random XOR/XNOR key gates into b01, b09, and \sasc{} and resynthesize them, producing six locks with random 8- and 16-bit keys (Table~\ref{tab:defense}); our results apply to this scheme. Across eight LOCA runs, all recovered RTL is proven equivalent to the locked circuits and retains the key inputs, but none recovers the original design. We also test eight additional locks on \texttt{usb\_phy}, two unpublished designs, and the \sasc{} mutant. Of nine Haiku runs, six produce proven RTL (two qualify as lifts), all retaining the key inputs. The other three fail: one worker returns no answer, and two runs are guard stops.

\begin{table}[t]

\caption{Defense results (successes/attempts). Obfuscation and sequences: Haiku; XOR/XNOR recovery: Haiku/Sonnet; key inference: GPT. \mbox{---: not} applicable; bold: all attempts succeeded.}

\label{tab:defense}
\centering\scriptsize
\setlength{\tabcolsep}{5pt}
\renewcommand{\arraystretch}{0.87}
\begin{tabular}{lccc}
\toprule
 & b01 & b09 & sasc\\
\midrule
\multicolumn{4}{l}{\emph{Obfuscation (function kept): LOCA lifted}}\\
Misleading names & 2/3 & 2/3 & \win{3/3}\\
Alt.\ resynthesis & 2/3 & \win{3/3} & \win{3/3}\\
Retiming & \win{3/3} & --- & \win{3/3}\\
Flip-flop duplication & 2/3 & \win{3/3} & \win{3/3}\\
State re-encoding & \win{3/3} & \win{3/3} & ---\\
\midrule
\multicolumn{4}{l}{\emph{XOR/XNOR locks: original design recovered}}\\
8-bit key & 0/1 & 0/2 & 0/1\\
16-bit key & 0/2 & 0/1 & 0/1\\
\midrule
\multicolumn{4}{l}{\emph{Correct keys guessed by GPT}}\\
8-bit key & 14/20 & 6/20 & 16/20\\
16-bit key & 18/20 & 4/20 & 11/20\\
\midrule
\multicolumn{4}{l}{\emph{FSM obfuscation: protected circuit lifted}}\\
4-step sequence & 2/3 & 0/3 & 2/3\\
8-step sequence & 0/3 & 1/3 & 2/3\\
\midrule
\multicolumn{4}{l}{\emph{Unlock sequence read by Haiku from the netlist}}\\
4-step sequence & \win{10/10} & \win{10/10} & 8/10\\
8-step sequence & \win{10/10} & \win{10/10} & 4/10\\
\bottomrule
\end{tabular}
\end{table}

We separately test whether LLMs can infer keys from locked netlists or lifted RTL. For the six locks in Table~\ref{tab:defense}, GPT guesses the correct key at least once for every lock, succeeding in 69/120 replies, compared with 6/150 for the Claude models. On b01's 16-bit lock, GPT succeeds in 9/10 replies using only the netlist. We verify these guesses against the original designs for evaluation, but an oracle-less attacker cannot confirm whether a guessed key is correct. On six further locks, GPT succeeds from the netlist alone in 27/80 replies on the unpublished designs and 26/40 on the \sasc{} mutant; Haiku succeeds in none of 60 replies.

We also test whether lifting helps infer keys. The results differ across lock sets. On four additional locks with proven Haiku LOCA outputs, GPT guesses correctly in 58/80 replies from lifted RTL versus 35/80 from the locked netlist. A predeclared replicate shows the same pattern (29/40 versus 14/40; stratified exact test, $p=0.001$). On the six public locks, however, the direction reverses: 30/60 from lifted RTL versus 39/60 from the netlist. We therefore draw no general conclusion about whether lifting helps key inference. Agreement among guesses is also unreliable: Haiku repeats the same wrong b09 key in all 10 RTL replies. Compactness may help rank candidate keys (Section~\mbox{\ref{sec:defense}-C}), but cannot confirm their correctness.

We next test sequential obfuscation in the style of HARPOON~\cite{HARPOON}, where an added finite-state machine (FSM) encodes an unlock sequence supplied through ordinary inputs after reset. Until the correct sequence arrives, the FSM holds the design in reset and corrupts some outputs. We protect b01, b09, and \sasc{} with 4- and 8-step sequences (Table~\ref{tab:defense}). Haiku with LOCA lifts the protected circuit in 7 of 18 runs. Of the other 11, three fail during cone extraction on one b09 variant, four are guard stops, and four produce proven RTL below the lifting threshold. These lifts preserve the protected circuit's behavior, including its unlocking conditions, which therefore appear in the recovered RTL.

We separately ask Haiku to identify the activation sequence encoded in the added FSM. It succeeds from the obfuscated netlist in 52 of 60 replies. A reply counts as correct if, after its sequence, the netlist is proven equivalent to the original design; a uniformly random sequence would succeed with probability at most 1/14 on one b09 variant and at most 1/254 on the others. Thus, limited success in lifting the protected circuit does not imply that its activation sequence stays hidden from an LLM attacker.

\subsection{Hidden cell functions}
We model camouflaging by giving each cell type a false label, used consistently across its instances. The attacker must infer the actual function behind each label; we call a candidate assignment a \emph{reading}. A proof establishes equivalence to the assumed reading; it does not establish that the reading matches the original design. On b01, even with knowledge of the defender's resynthesis script, filtering by random simulation leaves 379 possible readings.
We test the true reading and eight sampled alternatives using Haiku and Sonnet. The true reading produces the most compact RTL among these candidates, but only by a narrow margin. Repeated Haiku runs on the true reading and three close alternatives confirm this ranking. Sonnet also produces compact RTL from incorrect readings, so compression alone cannot confirm the true cell functions.

\emph{Takeaway for the defender.} Function-preserving obfuscation did not clearly reduce LOCA's recovery rate in our tests. With XOR/XNOR locking, LOCA recovers the keyed circuit, while recovering the original design requires the correct key. GPT often guessed that key, but the attacker cannot confirm it without a working chip or the original design. In the tested FSM protection, the netlist encodes an unlock sequence that Haiku identified in most replies. Hidden cell functions require guessing what each gate computes; equivalence to a guessed reading does not validate that guess, and compression distinguished the true reading from sampled alternatives only narrowly, by about the spread between repeated runs.


\section{Conclusion}

We presented LOCA, an LLM-based attack methodology for recovering readable, formally equivalent RTL from gate-level netlists. By dividing the netlist into complete next-state logic cones and verifying each rewrite, LOCA enables existing LLMs to recover designs that whole-netlist prompting cannot. It lifts 17 of 21 model--design pairs, compared with 6 for whole-netlist prompting, with final-version recovery up to 4,909 gates. Our ablation shows that smaller tasks alone are insufficient; functional boundaries and formal feedback both matter. In our tests, lifted RTL also helped an LLM answer questions about circuit behavior and make correct edits.

Defenses against LOCA in our experiments reveal a distinction between obscuring a netlist's structure and hiding its original function. Function-preserving obfuscation did not clearly reduce recovery in our tests, while the XOR/XNOR locks and hidden cell functions we tested prevented confirmed recovery of the original design. GPT correctly guessed some keys, but an attacker without a working chip cannot verify them. As LLM capabilities improve, our findings suggest that gate-level complexity alone may become a weaker barrier to hardware reverse engineering.


\section*{Content Generated by AI}
LLMs are the subject of this study. The authors also used Claude models (Anthropic) and OpenAI Codex as checkers and critics: they reviewed each experiment before it ran, cross-reviewed code, recomputed every reported number from the recorded runs, checked claims against the data, and critiqued drafts; they also assisted with coding and text editing. The authors designed the study, wrote the paper, and take full responsibility for it.

\label{maintextend}

\end{document}